\documentclass[conference]{IEEEtran}

\usepackage{graphicx}
\usepackage{epstopdf}
\usepackage[cmex10]{amsmath}
\usepackage{amsfonts}
\usepackage[ruled,linesnumbered]{algorithm2e}
\usepackage{algorithmic}
\usepackage{array}
\usepackage{eqparbox}
\usepackage{epsfig}
\usepackage{color}
\usepackage{graphicx}
\usepackage{times}
\usepackage{float}
\usepackage{stfloats}
\usepackage{graphicx}
\usepackage{subfigure}
\usepackage{type1cm}
\usepackage{url}
\usepackage{cite}
\usepackage{enumerate}
\usepackage{lineno}
\usepackage{amsthm}
\usepackage{amssymb}
\usepackage{bbm}
\usepackage{bm}
\usepackage[short]{optidef}

\DeclareMathOperator*{\argmin}{arg\,min}

\theoremstyle{definition}

\newcommand{\argmax}{\operatornamewithlimits{arg\,max}}

\begin{document}
%
\title{On Maximizing Sampling Time of RF-Harvesting Sensor Nodes Over Random Channel Gains}

\author{\IEEEauthorblockN{Changlin~Yang}
\IEEEauthorblockA{School of Computer Science\\
Zhongyuan University of Technology\\
Email: changlin@zut.edu.cn}
\and
\IEEEauthorblockN{Kwan-Wu~Chin}
\IEEEauthorblockA{School of Electrical, Computer and\\Telecommunications Engineering\\
University of Wollongong\\
Email: kwanwu@uow.edu.au}
\and
\IEEEauthorblockN{Ying Liu}
\IEEEauthorblockA{School of Electrical, Computer and\\Telecommunications Engineering\\
University of Wollongong\\
Email: yl694@uowmail.edu.au}
}
\maketitle

%
\begin{abstract}
In the future, sensor nodes or Internet of Things (IoTs) will be tasked with sampling the environment.  These nodes/devices are likely to be powered by a Hybrid Access Point (HAP) wirelessly, and may be programmed by the HAP with a {\em sampling time} to collect sensory data, carry out computation, and transmit sensed data to the HAP.  A key challenge, however, is random channel gains, which cause sensor nodes to receive varying amounts of Radio Frequency (RF) energy. %
To this end, we formulate a stochastic program to determine the charging time of the HAP and sampling time of sensor nodes.  Our objective is to minimize the {\em expected} penalty incurred when sensor nodes experience an energy shortfall.
We consider two cases: {\em single} and {\em multi} time slots.  In the former, we determine a suitable HAP charging time and nodes sampling time on a slot-by-slot basis whilst the latter considers the best charging and sampling time for use in the next $T$ slots.
We conduct experiments over channel gains drawn from the Gaussian, Rayleigh or Rician distribution.
Numerical results confirm our stochastic program can be used to compute good charging and sampling times that incur the minimum penalty over the said distributions. 
\end{abstract}
\begin{IEEEkeywords}
Wireless Sensor Networks, Wireless Charging, Stochastic Programs, Sample Average Approximation (SAA).
\end{IEEEkeywords}

%
\IEEEpeerreviewmaketitle

%
\section{\label{INTRO}Introduction}
Future buildings or cities will be instrumented with Internet of Things (IoTs) or low-power sensing devices \cite{Iot1}.  These devices will be tasked with collecting sensed data for a given period \cite{WSNSampling}.  They then report any sensed data to a Hybrid Access Point (HAP) or a sink.  As it is well known, these devices are energy constrained.  Hence, they must be able to harvest energy from the environment.  A promising source of energy is Radio Frequency (RF).  For example in \cite{WPT1}, the authors outline a sensor node prototype with a camera that harvests RF energy from Access Points (APs) transmissions; see \cite{RF1} and \cite{EHWPTsurvey1} for other examples.  Thus one can envisage an HAP first charging sensor nodes or IoTs via RF and tasking sensor nodes to switch on their sensor (camera) for a given period, process sensed data (images) and transmitting the result, e.g., whether an object is detected, back to the HAP. Alternatively, the HAP may program sensor nodes to send sensory data, e.g., temperature, continuously within their assigned sampling time.

\begin{figure}[htbp]
	\centering
	\includegraphics[scale=0.8]{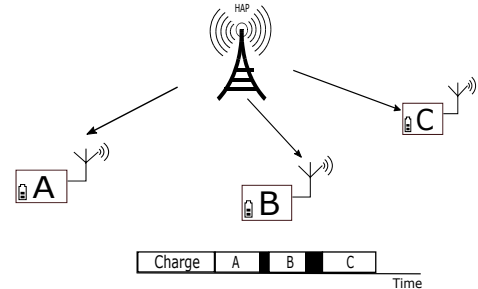}
	\caption{A example with an HAP and three sensor nodes; namely $A$, $B$ and $C$.  The HAP charges and programs the sensor nodes to be active for a given duration.  Black areas indicate idle times due to insufficient energy.}
	\label{FIG1}
\end{figure}

Figure \ref{FIG1} shows an HAP with three sensor nodes.  The HAP operates in a half-duplex manner and is responsible for charging these sensor nodes via RF and programming their respective sampling time.  Observe that if more time is afforded to charging, then there will be less time to collect samples from sensor nodes. Ideally, if the HAP has accurate channel gain information, each sensor node will have sufficient energy to utilize its allocated sampling period.  However, in practice, as the channel gain is {\em random}, the assigned sampling time of a sensor node is likely to be incorrect.  
In Figure \ref{FIG1}, the HAP has assigned the three sensor nodes a different sampling period.  However, node A and B are unable to fully utilize their allocated sampling time as they have exhausted their harvested energy; as indicated by the black areas.  Hence, node A and B are idle for some time.
On the other hand, node-C is able to fully utilize its active time.  Moreover, it may have residual energy.  Hence, a better schedule is to assign a shorter time period to node A and B and extends node C's sampling time.  In this example, we see the importance of assigning sampling times appropriately and keeping idle times to a minimum; this is our key {\em aim}.  The challenges are random channel gains and that collecting accurate channel gains incur energy that otherwise could be used for sampling and transmissions.  

Our problem resembles ``harvest-and-transmit'' works that aim to optimize the HAP charging time and also the transmission time of devices in Wireless Powered Communication Networks (WPCNs).  Their goal is to maximize the sum-rate or min rate of devices.  For example, reference \cite{TPUTWPN} and \cite{RFTDMA2} consider a HAP charging multiple RF-harvesting users.  The problem is to determine a charging time and transmission time of users that maximize the sum-rate or common throughput (min rate) of users.  The work in \cite{TPUTWPN} has been extended with an HAP with Multiple Input Multiple Output (MIMO) \cite{WPCNMIMO2}, full-duplex capabilities \cite{RFTDMA1} or relays \cite{RFNasir}. These works, however, assume perfect channel gains.  
To this end, in \cite{JBeamRF1}, the authors employ chance programming to ensure that the energy supplied to devices and their allocated time yields a certain data rate with a given probability.  On the other hand, reference \cite{NgNonL17} uses a robust optimization approach to ensure the derived charging and time allocation of devices remain valid for all channel gains within a given range.
In \cite{SWIPTLink16}, Liang et al. also consider maximizing sum rate.  In their problem, there are $N$ {\em fixed} size slots, and the problem is to determine the fraction of slots dedicated to uplinks and charging.  
In \cite{WXDSWIPT1}, Liu et al. aim to maximize the energy efficiency, which is defined as the number of bits received by the HAP over expended energy.  Advantageously, sensor nodes are able to harvest RF energy from each other's transmissions. 
In \cite{WPTBeam16},  Du et al. consider a base station that uses a sharp beam to transmit power.  They assume each sensor node has varying data arrival rates.  The problem is to determine a charging sequence that allows all nodes to have the longest operational lifetime. 
%
%

In the aforementioned works, except for \cite{JBeamRF1} and \cite{NgNonL17}, channel gains are assumed deterministic.  In \cite{JBeamRF1} and \cite{NgNonL17}, the authors assume one particular channel gain realization, either the worst case or one that ensures the solution is valid for a majority of possible channel gains.  That is, their solution is only optimized for a given channel gain value.  Consequently, the computed solution is likely to be conservative.
In contrast, we consider the cost or penalty of all channel gains realizations. Also, in these prior works, they only optimize over a single time slot.
For works that consider multiple slots, e.g., \cite{SWIPTLink16}, they assume the channel gain is known over multiple slots.  However, channel gains are likely to be even more inaccurate as compared to the single slot case.  
Prior ``harvest-and-transmit'' works assume nodes use {\em all} of their harvested energy.  This is reasonable as the quantity of interest is data rate, which is a function of the transmit power.  As explained later, in our case, nodes may have surplus energy that can then be used to cover any energy shortfall in subsequent time slots.
Lastly, the quantity of interest in prior works is throughput or sum-rate.  However, our focus is to minimize the expected penalty associated with idle slots.  

%
Henceforth, this paper makes the following {\em contributions}.  We consider the novel problem of setting the HAP charging time and sampling time of sensor nodes over random channel gains.  
We model the problem as a Stochastic Program (SP) and use it to maximize the minimum sampling time of each sensor node using the minimal HAP charging time.  Critically, the computed sampling time of nodes must minimize their expected idle times.  
With the SP in hand, we use it to study two cases: (i) {\em single slot}, where a new charging time and sampling time is computed for each slot, and (ii) {\em multi slots}, where the HAP and sensor nodes use respectively the same charging and sampling time over multiple time slots.  This case is significant because gathering accurate channel information is expensive.  Moreover, it reduces the need for the HAP to compute and communicate a new sampling time to sensor nodes in every slot. 
We solve the formulated SP using Sample Average Approximation (SAA).  We then use it to study the said cases over channel gains drawn from Gaussian, Rayleigh and Rician distributions.  Moreover, we compare our solution against approaches that either assume the best, average or worst (robust) channel gains.
Our results show that our SP is able to assign the best sampling time that ensures the minimum expected idle time that tends towards zero. 
They show that large variance in channel gains results in a significant longer idle time.  The results also show that solutions that assume the best, average or worst (robust) channel gains result in significant idle times.
Lastly, to the best of our knowledge, the application of SP to charging problems is new.  Hence, our paper may inspire its use as a solution to other interesting problems in WPCNs.
 
Next, in Section \ref{Preliminaries}, we introduce our network model, notations and provide a brief overview of SP. Following that, in Section \ref{Stochastic Programs}, we present our SP for both the single and multi-slot cases.  We outline SAA in Section \ref{SAA}.  Our evaluation methodology and results are presented in Section \ref{Evaluation}. The paper concludes in Section \ref{Conclusion}.

%
%
\section{Preliminaries}
\label{Preliminaries}
We consider a half-duplex HAP and a set $S$ of RF-harvesting devices, each indexed by $i$.  The HAP transmits with power $P$ (in Watts). 
Each device has an energy consumption rate of $P_c$, which includes the energy cost associated with sensing, processing and transmission.  Each device $i\in S$ is active for $T_i$ seconds.  Hence, when active, a device $i$ consumes $P_cT_i$ Joules of energy.  We assume devices inform the HAP of their battery level and the channel gain in the previous slot.  Devices have a half-duplex radio where in time $\tau$ it harvests energy, and in time $T_i=1-\tau$, it is allowed to transmit data to the HAP. 
%

Devices have a super capacitor with available energy $B_i$ and a maximum capacity of $\mathcal{B}$.  The capacitor $B_i$ is initially empty and is replenished by the HAP thereafter.  
Each device $i$ has an RF energy harvester, e.g., \cite{WPT1}, with energy conversion efficiency of $\eta$.  Let $g_i$ be the channel gain, a random variable, from the HAP to sensor node $i$; as we will see in Section \ref{Evaluation}, $g_i$ follows either a Gaussian, Rayleigh or Rician distribution.  
%
%
%
%
The amount of RF energy that is harvested from the HAP is,
\begin{equation}
E_i = \eta P g_i \tau
\label{ENERGYEQU}
\end{equation}
where $\tau$ is the charging time used by the HAP.

We now present a brief description of two-stage stochastic programs; see \cite{StocProgBook} for more details.
A two-stage stochastic program has the following structure,
\begin{mini}
  {x}{\label{SProgEX1}x + \mathbb{E}_{\omega}[h(x, \omega)]}{}{}
  \addConstraint{Ax}{\ge b,}{}
  \addConstraint{x}{\ge 0}{}
\end{mini}
The goal is to select the minimal $x$ value that also minimizes the expected value of $h(.,.)$; aka the second-stage problem.  Note, the expectation is taken with respect to the random variable $\omega$.  The second-stage problem is,
\begin{mini}
  {y}{\label{SProjEX2}g_{\omega}y}{}{h(x,\omega)=}
  \addConstraint{W_\omega y}{\ge r_\omega - T_\omega x,}{}
  \addConstraint{y}{\ge 0}{}
\end{mini}
where $W_w y$ is the recourse taken to ensure the constraint remains feasible given the $x$ value from the first stage SP. 
As an example, assume a problem instance has two realizations of $T_\omega$, say $\{2, 5\}$; each of which occurs with probability 0.5, and $g_\omega=W_\omega=r_\omega=1$. Then we can write the following deterministic equivalent LP,
\begin{mini}
  {x}{\label{SProgEX1a}x + 0.5y_1 + 0.5y_2}{}{}
  \addConstraint{Ax}{\ge b,}{}
  \addConstraint{y_1}{\ge 1-2x}{}
  \addConstraint{y_2}{\ge 1-5x}{}
  \addConstraint{x,y_1,y_2}{\ge 0}{}
\end{mini}
Observe that a copy of (\ref{SProjEX2}) is made for every realization and the resulting problem (\ref{SProgEX1a}) is a deterministic LP.
A key challenge is that there can be infinitely many scenarios or realizations.  Hence, the resulting {\em equivalent LP} becomes very large; the key topic addressed in Section \ref{SAA}.  
Lastly, given a $\omega$, the decision variable $y$ in (\ref{SProjEX2}) is the {\em recourse} action taken given $x$.  That is, if the given $x$ value from the first-stage problem is insufficient to satisfy the constraints in the second stage problem once $\omega$ is known, then $y$ needs to take on a non-negative value.

\section{Stochastic Programs}
\label{Stochastic Programs}
We will first present a two-stage stochastic program to determine the charging and sampling time of sensor nodes or devices in a single slot.  After that, we extend the formulation to consider charging and sampling times that are used repeatedly over $T$ slots.
\subsection{Single Slot}
Our problem is to maximize the minimum active time $T_i$, i.e.,  $\text{MAX}\;\text{MIN} \{T_i\}_{i\in S}$, using the minimal charging time $\tau$.  As mentioned, the channel gain $g_i$ is a random variable.  Consequently, after the HAP charges devices for $\tau$ time and programs device $i$ to collect and transmit samples for $T_i$ time, a device $i$ may find that its harvested RF energy is insufficient; i.e., $T_iP_a> E_i$.  When this happens, device $i$ becomes idle.

We now formulate a stochastic program for the problem at hand.  It has the following first stage problem,
\begin{maxi}
  {\tau, \Delta}{\label{SProgEX1c} \text{MIN}\{T_i\} - \mathbb{E}[h(\tau, \Delta, \bm{g})]}{}{}
  \addConstraint{\tau + \sum_{i\in S} T_i}{=1}{}
\end{maxi}
where $\bm{g}=\{g_i\}_{i\in S}$ and $\Delta=\{T_i\}_{i\in S}$. The expectation is taken with respect to the joint probability distribution of all channel gains from the HAP to each device.
Note, in order to solve (\ref{SProgEX1c}), a standard trick is to replace $\text{MIN}\{T_i\}$ with $Z$ and bound the value of $Z$ to be less than equal to the smallest $T_i$ value; i.e., we have the following constraint,
\begin{align}
Z \le T_i, &\;\forall i\in S
\label{SPC3}
\end{align}
The second-stage problem is defined as,
\begin{equation}
h(\tau, \Delta, \bm{g}) = \text{MIN} \sum_{i \in S} w_iy_i
\end{equation}
subject to
\begin{align}
& P_{c}y_i + B_i - P_{c}T_{i} + E_i \ge 0, & \quad \forall i \in S \label{SP2_1}\\
& y_i \ge 0, & \quad \forall i \in S \label{SP2_3}
\end{align}
The decision variable $y_i$, aka recourse, with penalty $w_i$, corresponds to the idle time of device $i$.  This means for $y_i$ time, device $i$ has no energy. 
In the sequel, we define a node to have {\em surplus energy} if after subtracting its consumed energy, the node's remaining energy is positive. 
Observe that if $\tau$ is sufficient then a node has surplus energy and in constraint (\ref{SP2_1}), $y_i$ can be set to zero.
\subsection{Multiple Slots}
The previous formulation considers maximizing the minimum $T_i$ value over one time unit/slot or period.  In this section, we give a formulation that allows the HAP to determine a $\tau$ and the $T_i$ value for $T$ slots.  That is, for each $t=1,\ldots,T$, the same $\tau$ and $T_i$ value is used by the HAP and devices, respectively. 
In the sequel, we use the superscript `ts' to denote variables and coefficients that exist in slot $t$ of scenario $s$.  
A scenario or sample path $s$ is denoted as $\mathbf{g}^{s}=\{\mathbf{g}^{ts}\}_{t\in T}$; e.g., if there are two nodes $\{A,B\}$ and $T=3$, then we have $\mathbf{g}^{s}=\{\{g^{1s}_A, g^{1s}_B\}, \{g^{2s}_A, g^{2s}_B\}, \{g^{s3}_A, g^{3s}_B\}\}$.  Let $\Omega$ be a set containing all possible scenarios.

We have the following two-stage stochastic problem,
\begin{maxi}
  {\tau, \Delta}{\label{SProgMultiEX1c} \mathbb{E}_{\Omega}\left[h(\tau, \Delta, \mathbf{g}^s)\right] }{}{}
  \addConstraint{\tau + \sum_{i\in S} T_i}{=1, }{\;\forall i\in S}
\end{maxi}
where $h(\tau, \Delta, \mathbf{g}^s)$ is an LP for scenario $s$,
\begin{equation}
 h(\tau, \Delta, \mathbf{g}^s) = MAX \sum_{t=1}^T Z^{ts} - \sum_{i\in S} y^{ts}_i
 \label{multi2}
\end{equation}
subject to,
\begin{align}
P_c y_i^{ts} + B_i^{(t-1)s} - P_{c}T_{i} + \eta Pg^{ts}_i\tau & \ge 0, \label{multi3} \\ 
%
P_c y_i^{ts} + B_i^{(t-1)s} - P_{c}T_{i} + \eta Pg^{ts}_i\tau & = B_i^{ts}+\delta_i^{ts}, \label{multi4}\\ 
Z^{ts} & \le T_i,
\end{align}
where the aforementioned constraints are iterated over all device $i\in S$ and $t=1,\ldots, T$. 
Note, $\delta_i^{ts}$ denotes the excess energy that is loss after node $i$'s battery is full. 

%
Observe that the problem is to determine a suitable $\tau$ and $T_i$ value to be used in all scenarios over $T$ slots such that the sum of the minimum active period $T_i$ minus the time due to idle slots is maximized.  
%


\subsection{Discussion}
The key challenge when solving the aforementioned stochastic programs is the number of scenarios; i.e., there are exponentially many instances of $\bm{g}$.  Consider the single-slot case.  If each node has $b$ channel gain levels, then there will be $b^{|N|}$ scenarios, meaning the equivalent LP becomes computationally intractable with increasing number of devices or channel gain levels. 
The multi-slots case is even more challenging because $\Omega$ contains orders of magnitude more scenarios than the single-slot case.
Next, we present the approach used to obtain a solution to both stochastic programs.
\section{Sample Average Approximation (SAA) \label{SAA}}
SAA is a Monte Carlo simulation approach to solving stochastic programs \cite{StocProgBook}. 
The basic idea is to replace the expectation term of (\ref{SProgEX1c}) with its sample mean approximation,
\begin{equation}
\mathbb{E}[h(\tau, \Delta, g_i^t)] = \frac{1}{N}\sum_{j=1}^{N} [h(\tau, \Delta, \bm{g^j})]
\label{SAA1}
\end{equation}
where $N$ is the number of sampled scenarios, and $\bm{g^j}$ is a channel gain realization of all $|S|$ devices; each element is an individual independent distribution (i.i.d) sample. 

Let $z_N$ denote the objective value of (\ref{SProgEX1c}) computed using SAA with $N$ scenarios, and the corresponding solution is $\hat{\tau}$.
To obtain a sample average of $z_N$, denoted as $\bar{z}_N$, we solve problem (\ref{SProgEX1c}) $M$ times, each with $N$ different scenarios, where $\mathbb{E}[h(\tau, \Delta, g_i^t)]$ of (\ref{SProgEX1c}) is estimated using (\ref{SAA1}).  We thus have $z_N^1, \ldots, z_N^M$, each with corresponding solution $\hat{\tau}^1, \ldots\hat{\tau}^M$.  Then, we have,
\begin{equation}
\bar{z}_N = \frac{1}{M} \sum_{m=1}^M z^m_N
\end{equation}

For each solution $\hat{\tau}$, we can estimate its objective value as,
\begin{equation}
\hat{z}_{N'}(\hat{\tau}) = \hat{\tau} + \frac{1}{N'} \sum_{j=1}^{N'} h(\hat{\tau}, \Delta, \bm{g}^j)
\label{Equ18}
\end{equation}
where $N'\gg N$, and $N'$ of the $\bm{g}^j$ samples are independent from those used to compute $z_N$.
%
The optimal solution is then
\begin{equation}
\hat{\tau}^* \in\argmax\{ \hat{z}_{N'}(\hat{\tau})\;|\;\hat{\tau}\in\{\hat{\tau}^1,\ldots, \hat{\tau}^M\} \}
\end{equation}

%
One can measure the quality of a solution, namely $\hat{\tau}^*$ and $\Delta^*$, via the gap $\hat{z}_N(\hat{\tau}^*)-\bar{z}_N$.  Its estimated variance is,
\begin{equation}
\hat{\sigma}^2_{\text{gap}} = \hat{\sigma}^2_{\hat{z}_{N'}(\hat{\tau}^*)} + \hat{\sigma}^2_{\bar{z}_N}
\label{estVariance}
\end{equation}
where $\hat{\sigma}^2_{\hat{z}_{N'}(\hat{\tau}^*)}$ is defined as
%
\begin{equation}
\frac{1}{(N'-1)N'}\sum^{N'}_{t=1}\left(\hat{\tau}^*+ h(\hat{\tau}^*, \Delta^*, \bm{g}^t)-\hat{z}_{N'}(\hat{\tau}^*) \right)^2
\end{equation}
and $\hat{\sigma}^2_{\bar{z}_N}$ is,
\begin{equation}
\frac{1}{(M-1)M}\sum^{M}_{m=1}\left(z^m_N - \bar{z}_N\right)^2
\end{equation}
In our experiments, we use an $M$ and $N$ value that ensures the variance of the gap $\hat{\sigma}^2_{\text{gap}}$ is less than $10^{-3}$. 

\section{Evaluation}
\label{Evaluation}
We now use our formulated SP with SAA approach, abbreviated as {\em SpSaa}, to study various scenarios.  We set our parameter values according to current systems. Specifically, the sink's transmit power is 200 mW, as per the 2.4 GHz 802.11b WiFi router in \cite{RFPowersim}, equipped with an antenna that has a gain of 3 dBi. The power consumption rate of devices is set to 50 mW; similar to a Waspmote \cite{sim1}. All nodes are equipped with a super capacitor, which is initially empty and has a maximum capacity of 100mA. 

\begin{figure}[htbp]
	\includegraphics[scale = 0.4]{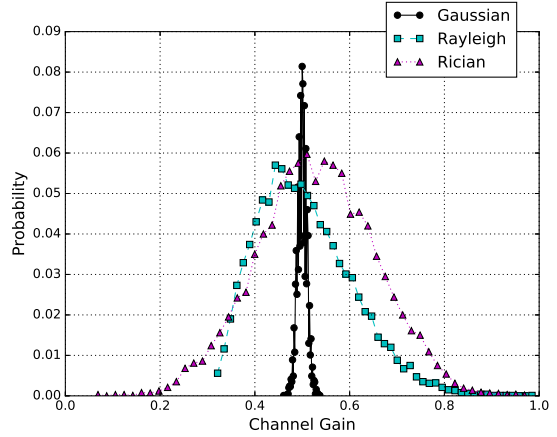}
  \caption{The probability distribution of Gaussian, Rayleigh and Rician. }
  \label{result0}
\end{figure}

\begin{figure*}[htbp]
  \centering
   \mbox{
    \subfigure[ \label{result1_1}]{\includegraphics[scale = 0.32]{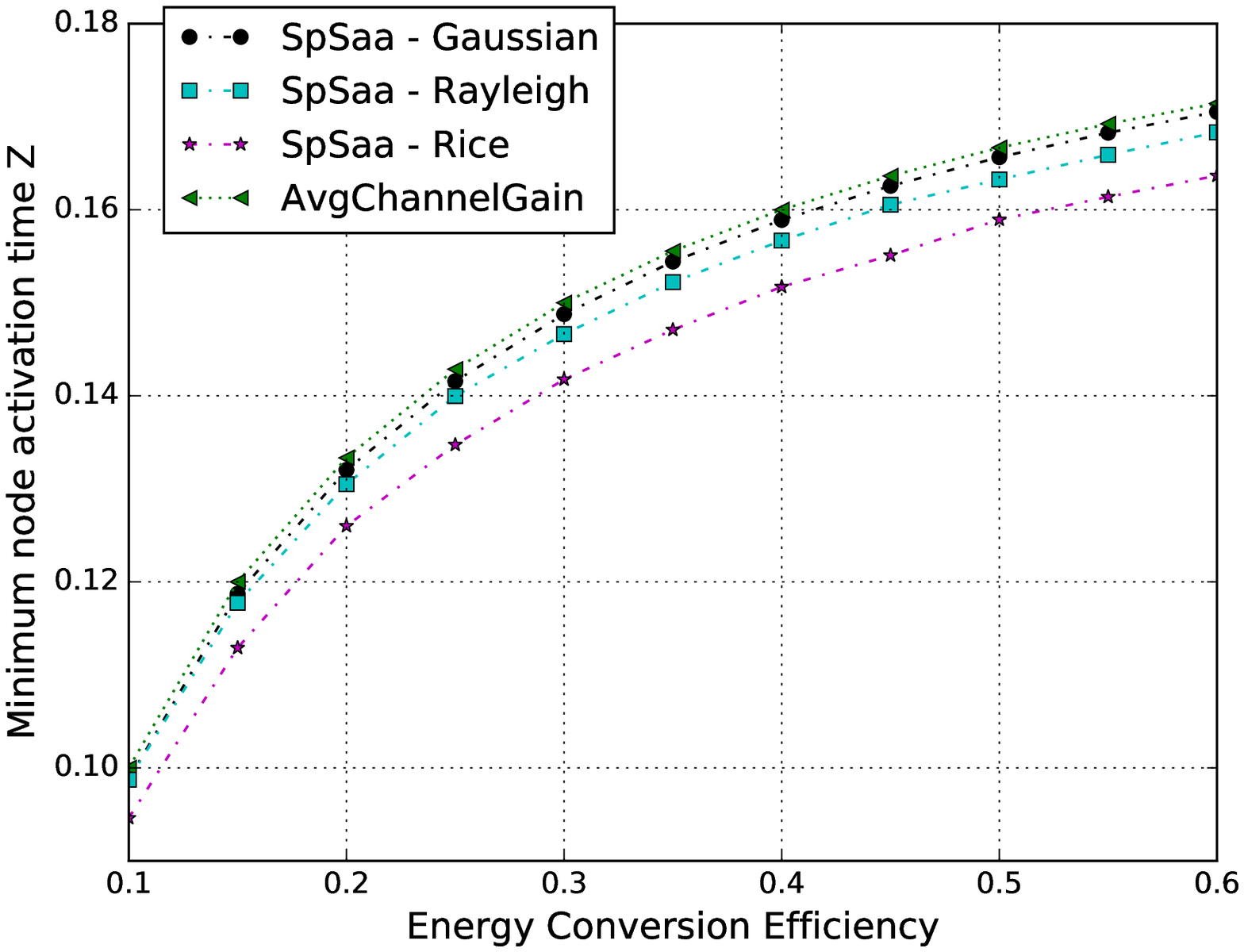}}\hspace*{-2.0em}
     \subfigure[ \label{result1_2}]{\includegraphics[scale = 0.32]{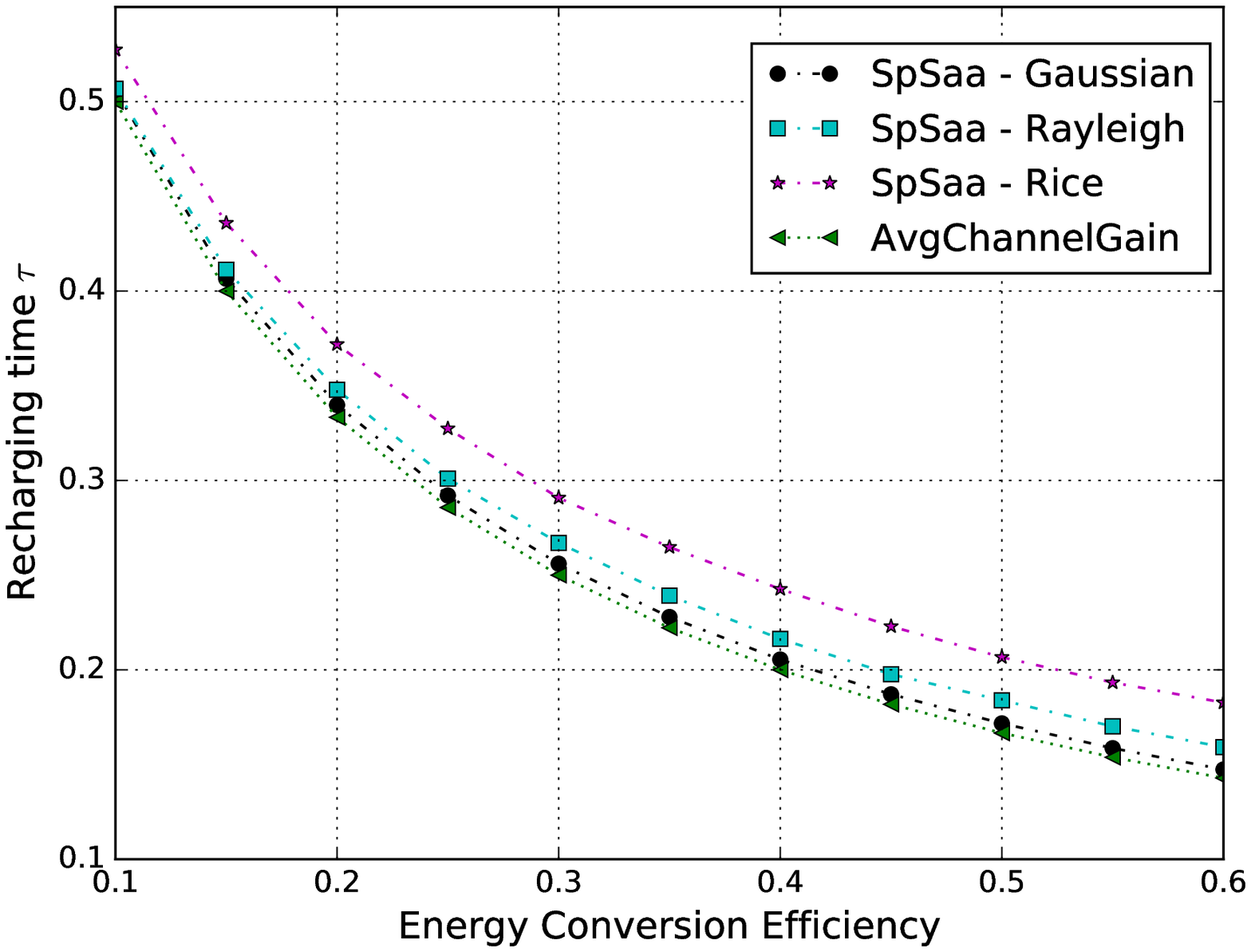}}\hspace*{-2.0em}
     \subfigure[ \label{result1_3}]{\includegraphics[scale = 0.32]{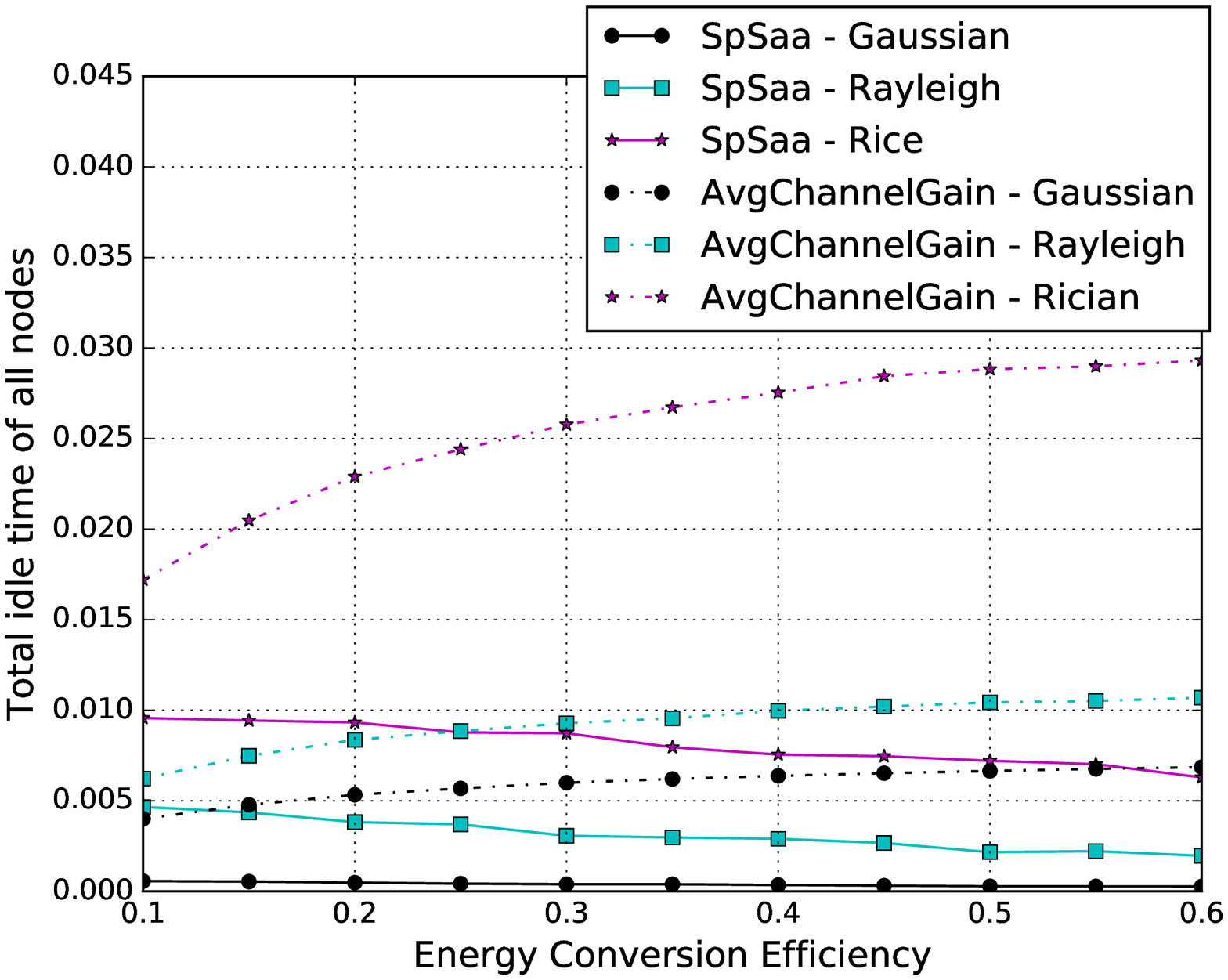}}
   }
  \caption{The slot average (a) idle time, (b) maximum $Z$, and (c) recharging time with increasing energy conversion efficiency}
  \label{ThreeFigures2}
\end{figure*}

We conduct two set of experiments; namely, single and multi-slots. In both experiments, five sensor nodes are randomly deployed on a given sensing field. The channel gain from the HAP to each node is generated according to one of the following distributions (see Figure \ref{result0}): Gaussian, Rayleigh and Rician. The Gaussian distribution has a variance of 0.1, the Rayleigh has a scale parameter of two, and Rician has an non-centrality parameter of four. All these three distributions are scaled to be within $[0,1]$ and have a mean of 0.5.
In single slot experiments, we verify the performance of SpSaa where scenarios are generated as per the said distributions. We also compare SpSaa against the scenario where the channel gains take on their average value, i.e., 0.5.  Note, this case is usually the practice in past works whereby only the average or fixed channel gains are used to compute a solution.
In multi-slot experiments, we only report results for the Rician distribution as the other two distributions exhibit similar results.  
We compare SpSaa against the worst, average and best channel gain scenarios. Specifically, in these scenarios, the channel gain is respectively 0.01, 0.5 and 1.0. In the sequel, we label these scenarios as MinChannelGain, AvgChannelGain and MaxChannelGain, respectively.

In multi-slots experiments, we use the suffix `Single', for instance, `SpSaaSingle', to indicate the case when SpSaa runs in a {\em slot-by-slot} manner. Specifically, after the algorithms derive the recharging time $\tau$ and corresponding node activation schedule $T_i$, all nodes update their battery level based the actual channel gain.  They then send their updated battery level to the sink. In the next slot, the algorithms then recalculate a new recharging time $\tau$ and node activation schedule. 

The suffix `Multi' indicates multi-slots scenarios, where each scenario consists of a sequence of possible channel gain value in each of the $T$ slots.  For example, SpSaaMulti considers all possible scenarios that may occur over $T$ slots.  It then calculates the best recharging time and node activation schedule to be used in each slot for the next $T$ slots.  

We record the following metrics: max-min node activation time $Z$, recharging time $\tau$ and idle time, of all nodes. Note, in order to calculate the idle time for MinChannelGain, AvgChannelGain and MaxChannelGain scenarios, we substitute their calculated $Z$ and $T_i$ value into Equ. (\ref{Equ18}). 

\subsection{Single Slot}
%
%
%
We first study varying energy conversion efficiency.  As expected, from Figure \ref{result1_1}, the minimum node activation time $Z$ is higher with better energy conversion efficiency. For the same reason, see Figure \ref{result1_2}, the HAP is able to reduce its recharging time.  Figure \ref{result1_1} also shows that the minimum node active time under the Gaussian distribution model when $\eta = 0.6$ is close to AvgChannelGain and higher than under the Rayleigh and Rice distribution model by 0.005 and 0.007 seconds, respectively.  
This is because the Gaussian distribution has the lowest variance.  Hence, when SpSaa runs with the same $M$ and $N$ values, the gap to the optimal SP is significantly smaller; see Equ. (\ref{estVariance}).  For the same reason, as shown in Figure \ref{result1_3}, SpSaa under the Gaussian model has the smallest idle time, followed by SpSaa under the Rayleigh model with an idle time that is less than 0.002 seconds at $\eta=0.6$.  In contrast, under the Rician model, SpSaa has the maximum idle time of 0.007 seconds when $\eta =0.6$. 

\begin{figure}[htbp]
	\includegraphics[scale = 0.4]{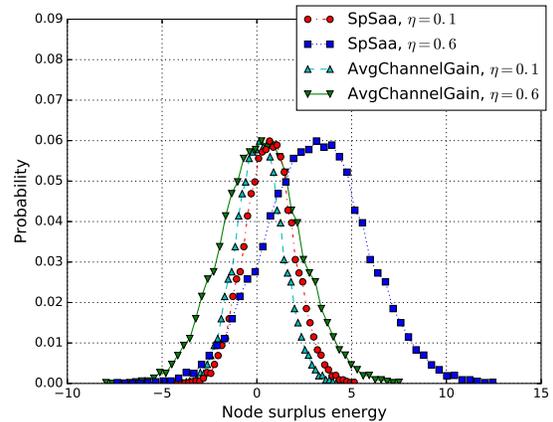}
  \caption{The probability distribution of surplus energy at nodes when conversion efficiency equals 0.1 and 0.6 under the Rician model. }
  \label{result1_add}
\end{figure}

From Figure \ref{result1_3}, we observe that the total idle time of SpSaa under the Rayleigh and Rician models reduces with increasing conversion efficiency. 
This is because of the higher energy conversion efficiency resulting in more surplus energy at nodes.  For example, when $\eta=0.1$, the recharging time $\tau$ and minimum node active time $Z$ of SpSaa under the Rician model is respectively 0.527 and 0.095 seconds; when $\eta=0.6$, these two values are 0.183 and 0.163 seconds, respectively.  Given the aforementioned recharging time and minimum node active time, we can calculate the probability distribution of a node's surplus energy using Equ. (\ref{SP2_1}).  This probability is shown in Figure \ref{result1_add}, for energy conversion efficiency of 0.1 and 0.6 under the Rician model. 
We can see that when using SpSaa, the probability that a node has positive surplus energy is much higher when conversion efficiency equals 0.6.  Therefore, there are fewer recourses with increasing energy conversion efficiency.  This equates to less idle times. 
Figure \ref{result1_add} shows the variance of surplus energy calculated by AvgChannelGain under Rician model increased with a higher energy conversion efficiency, and its mean constantly equals zero. Therefore, although the probability that a scenario requires recourse is the same when conversion efficiency equals 0.1 and 0.6, the total number of recourses is higher for the latter case.  This explains why in Figure \ref{result1_3} the total idle time of AvgChannelGain under Gaussian, Rayleigh and Rician models is higher with increasing energy conversion efficiency. 

\subsection{Multiple Slots}
\begin{figure*}[htbp]
  \centering
   \mbox{
    \subfigure[ \label{result3_1}]{\includegraphics[scale = 0.32]{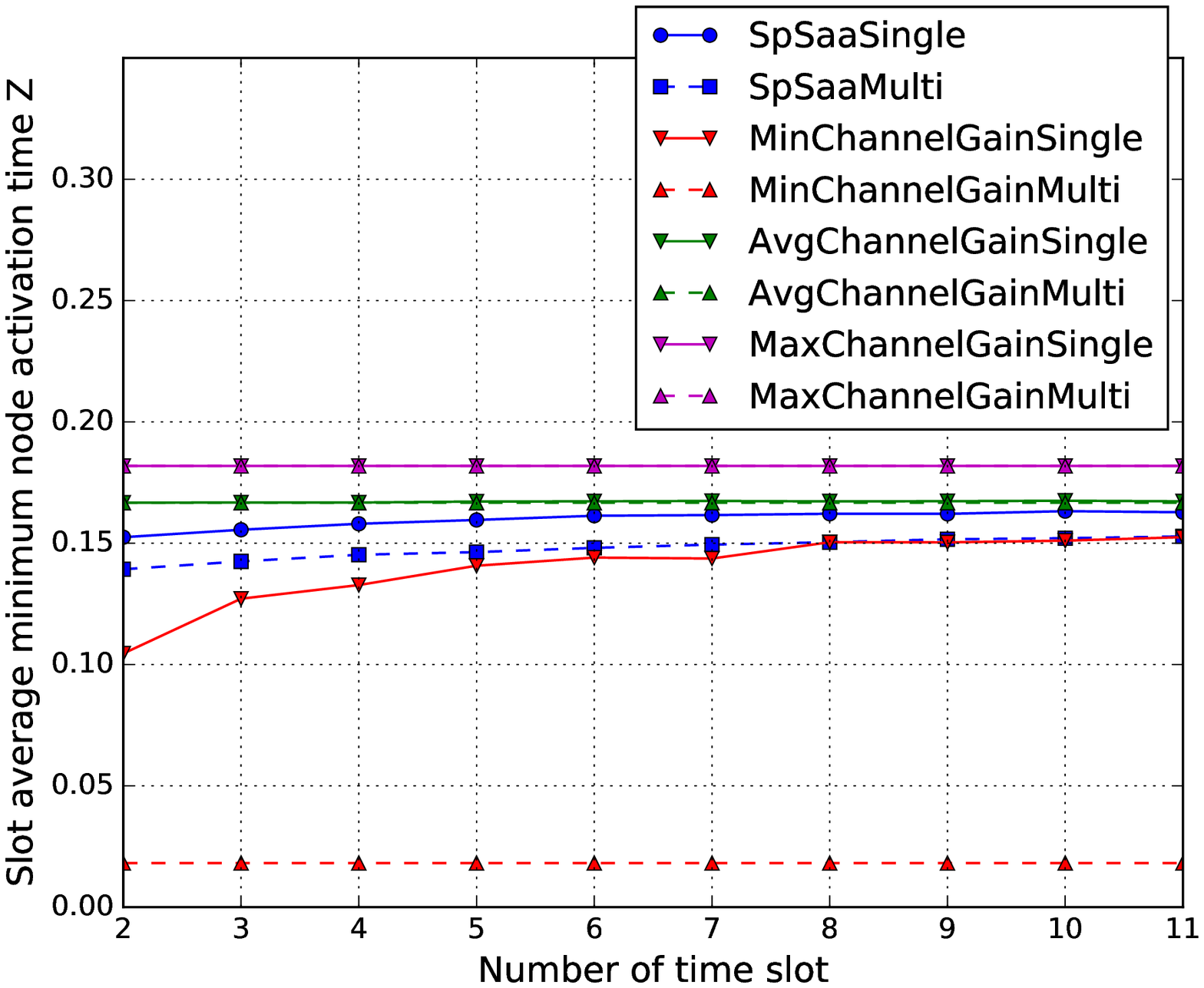}}\hspace*{-2.0em}
     \subfigure[ \label{result3_2}]{\includegraphics[scale = 0.32]{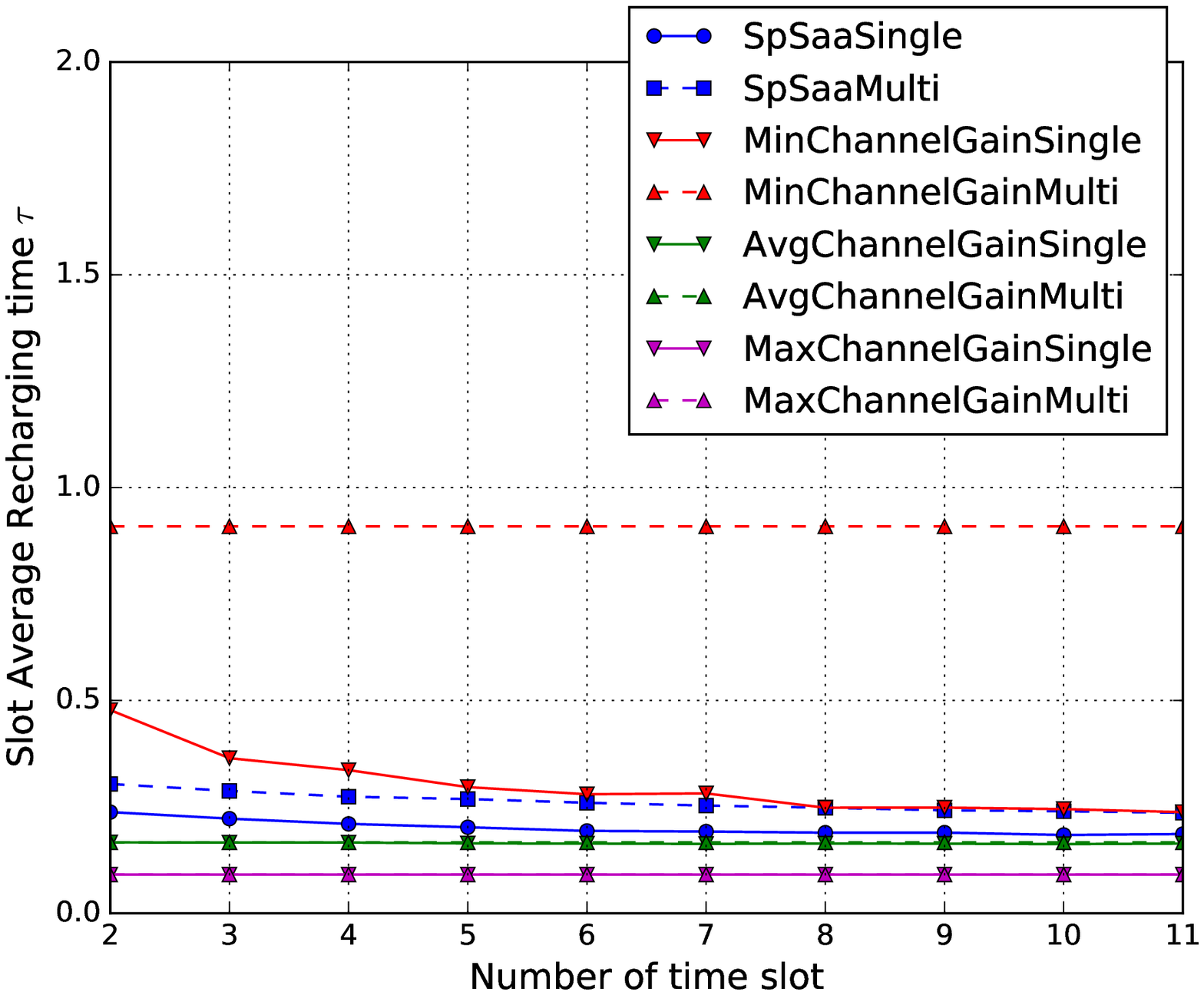}}\hspace*{-2.0em}
     \subfigure[ \label{result3_3}]{\includegraphics[scale = 0.32]{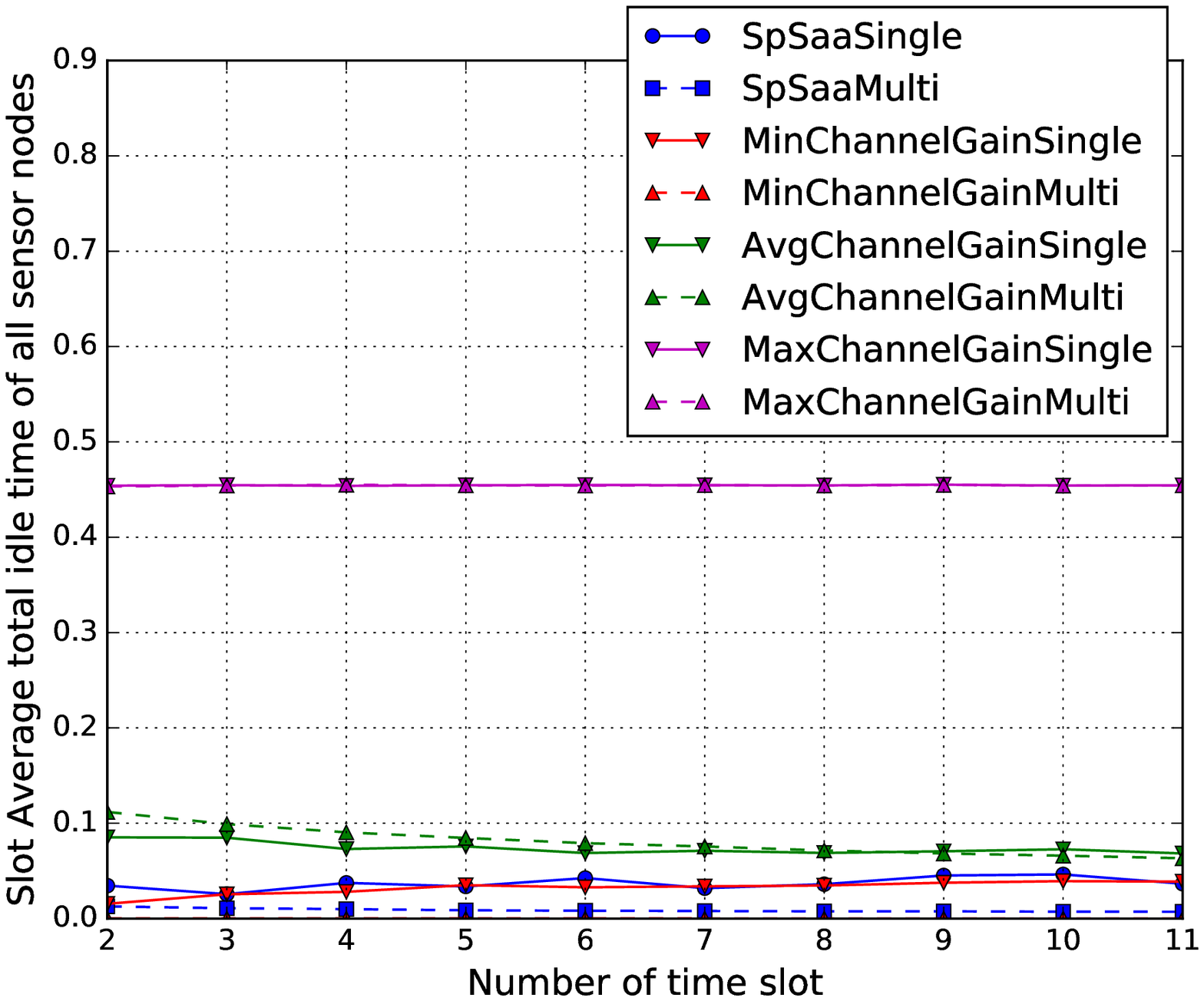}}
   }
  \caption{Time average of (a) idle time, (b) maximum $Z$, and (c) recharging time.}
  \label{ThreeFigures2a}
\end{figure*}

In this experiment, we fix the energy conversion efficiency to 0.4 and vary the number of time slots from two to 11 with an interval of one. 
From Figure \ref{result3_1}, we see that the minimum node activation time of MinChannelGainSingle dramatically increased from 0.1 to 0.15 seconds, whilst that of MinChannelGainMulti remained at 0.02 seconds.  From Figure \ref{result3_2}, the recharging time of MinChannelGainSingle reduced by half from 0.5 to 0.25 seconds whilst MinChannelGainMulti has a recharging time that exceeds 0.9 seconds. The reason is because when the HAP assumes the minimum channel gains, it uses a conservatively high recharging time.  However, the actual channel gain is almost alway higher than the minimum channel gain. 
Also, in contrast to the single slot case, where nodes report their actual channel gain to the HAP after each slot, the HAP knows the battery level of nodes.  This allows the HAP to consider any surplus energy in the next time slot. This results in a longer active time and shorter recharging time. 


From Figure \ref{result3_3}, we see that both MaxChannelGainMulti and MaxChannelGainSingle have a very long idle time at 0.45 seconds. This is because it only considers the maximum possible channel gain scenario.  This means the HAP assigns a very short recharging time.  This, however, may not be suitable for a majority of the scenarios.
Figure \ref{result3_3} also shows that SpSaaSingle has a longer idle time than SpSaaMulti. This is because in single slot runs, HAP updates the battery of all nodes in every slot according to the actual channel gain reported by sensor nodes.  However, this channel gain information is only for one scenario and channel gains are i.i.d across slots. Therefore, as compared to SpSaaMulti, SpSaaSingle greedily assigns less recharging time to achieve a longer node activation time; see Figure \ref{result3_1} and \ref{result3_2}. As a result, the minimum node activation time of SpSaaSingle is higher than SpSaaMulti for 0.01 seconds, however, its idle time is 0.05 seconds longer. 
%


%
From Figure \ref{result3_3}, we also see that the idle time of AvgChannelGainSingle and AvgChannelGainMulti reduces with increasing number of time slots. 
This is because the HAP uses a constant recharging time of 0.167 seconds, as shown in Figure \ref{result3_2}, and nodes have a constant minimum node activation time of 0.167 seconds; see Figure \ref{result3_1}. Consequently, nodes consume 8.33 Joules and harvest $16.67 \times g^t_i$ Joules energy per slot. This means when a node has a channel gain higher than 0.5, it has surplus energy that can be used in subsequent slots when there is energy shortfall. 

From the foregone experiments, SpSaaMulti uses 0.05 seconds less recharging time and 0.01 seconds less minimum node activation time as compared to SpSaaSingle.  However, the per slot total idle time of all sensor nodes of SpSaaSingle is higher than SpSaaMulti for 0.05 seconds. It is worth noting that the idle time calculated by SpSaaMulti is negligible. Moreover, SpSaaMulti does not require nodes to report their battery level at each time slot. 
\section{Conclusion}
\label{Conclusion}
This paper has studied the following novel problem: setting the HAP's charging and sampling times of nodes over random channel gains.  The goal is to ensure sensor nodes experience minimal idle times.  Moreover, we consider the problem in single and multiple slots scenarios. The problem is challenging due to the number of scenarios or channel gain realizations that increases with the number of nodes and time slots.  We use SAA to solve our stochastic program and use it to study the problem over channel gains drawn from three different probability distributions. Simulation results show that our approach ensures a minimum idle time and good sampling times for all tested distributions. 

\bibliographystyle{IEEEtran}
\bibliography{ref}

\begin{thebibliography}{10}
\providecommand{\url}[1]{#1}
\csname url@samestyle\endcsname
\providecommand{\newblock}{\relax}
\providecommand{\bibinfo}[2]{#2}
\providecommand{\BIBentrySTDinterwordspacing}{\spaceskip=0pt\relax}
\providecommand{\BIBentryALTinterwordstretchfactor}{4}
\providecommand{\BIBentryALTinterwordspacing}{\spaceskip=\fontdimen2\font plus
\BIBentryALTinterwordstretchfactor\fontdimen3\font minus
  \fontdimen4\font\relax}
\providecommand{\BIBforeignlanguage}[2]{{%
\expandafter\ifx\csname l@#1\endcsname\relax
\typeout{** WARNING: IEEEtran.bst: No hyphenation pattern has been}%
\typeout{** loaded for the language `#1'. Using the pattern for}%
\typeout{** the default language instead.}%
\else
\language=\csname l@#1\endcsname
\fi
#2}}
\providecommand{\BIBdecl}{\relax}
\BIBdecl

\bibitem{Iot1}
A.~Zanella, N.~Bui, A.~Castellani, L.~Vangelista, and M.~Zorzi, ``Internet of
  things for smart cities,'' \emph{IEEE Internet of Things Journal}, vol.~1,
  no.~1, pp. 22--32, 2014.

\bibitem{WSNSampling}
S.~Li, D.~X. Li, and X.~Wang, ``Compressed sensing signal and data acquisition
  in wireless sensor networks and internet of things,'' \emph{IEEE Transactions
  on Industrial Informatics}, vol.~9, no.~4, pp. 2177 -- 2186, 2013.

\bibitem{WPT1}
V.~Talla, B.~Kellogg, B.~Ransford, and S.~Naderiparizi, ``Powering the next
  billion devices with {Wi-Fi},'' in \emph{ACM CoNEXT}, Germany, Dec. 2015.

\bibitem{RF1}
X.~Lu, P.~Wang, D.~Niyato, D.-I. Kim, and Z.~Han, ``Wireless networks with {RF}
  energy harvesting: A contemporary survey,'' \emph{IEEE Communication Surveys
  and tutorials}, vol.~17, no.~2, pp. 757--767, 2015.

\bibitem{EHWPTsurvey1}
N.~Bhatti, M.~Alizai, A.~Syd, and L.~Mottola, ``Energy harvesting and wireless
  transfer in sensor network applications: Concepts and experiences,''
  \emph{ACM Transactions on Sensor Networks}, vol.~12, no.~3, p.~24, 2016.

\bibitem{TPUTWPN}
H.~Ju and R.~Zhang, ``Throughput maximization for wireless powered
  communications networks,'' \emph{IEEE Trans. on Wireless Communications},
  vol.~13, no.~1, pp. 418--428, Jan. 2014.

\bibitem{RFTDMA2}
Z.~Hadzi-Velkov, I.~Nikoloska, G.~K. Karagiannidis, and T.~Q. Duong, ``Wireless
  network with energy harvesting and power transfer: Joint power and time
  allocation,'' \emph{IEEE Signal Processing Letters}, vol.~23, no.~1, pp.
  50--54, 2016.

\bibitem{WPCNMIMO2}
L.~Liu, R.~Zhang, and K.-C. Chua, ``Multi-antenna wireless powered
  communication with energy beamforming,'' \emph{IEEE Transactions on
  Communications}, vol.~62, no.~12, pp. 4349--4352, Dec. 2014.

\bibitem{RFTDMA1}
X.~Kang, C.~K. Ho, and S.~Sun, ``Optimal time allocation for dynamic
  {TDMA}-based wireless powered communication networks,'' in \emph{IEEE
  GLOBECOM}, Austin, USA, Dec. 2014.

\bibitem{RFNasir}
A.~A. Nasir, X.~Zhou, S.~Durrani, and R.~A. Kennedy, ``Relaying protocols for
  wireless energy harvesting and information processing,'' \emph{IEEE Trans. on
  Wireless Comms}, vol.~12, no.~7, pp. 3622--3636, 2013.

\bibitem{JBeamRF1}
Q.~Sun, G.~Zhu, C.~Shen, X.~Li, and Z.~Zhong, ``Joint beamforming design and
  time allocation for wireless powered communication networks,'' \emph{IEEE
  Communications Letters}, vol.~18, no.~10, pp. 1783--1787, 2014.

\bibitem{NgNonL17}
E.~Boshkovska, D.~W.~K. Ng, M.~Zlatanov, A.~Koelpin, and R.~Schober, ``Robust
  resource allocation for {MIMO} wireless powered communication networks based
  on a non-linear eh model,'' \emph{IEEE Transactions on Communications},
  vol.~65, no.~5, pp. 1984--1999, 2017.

\bibitem{SWIPTLink16}
Z.~Liang and Y.~Liu, ``Link scheduling in {SWIPT} systems,'' in \emph{IEEE
  Globecom}, Washington DC, USA, Dec. 2016.

\bibitem{WXDSWIPT1}
T.~Liu, X.~Wang, and L.~Zheng, ``A cooperative {SWIPT} scheme for wirelessly
  powered sensor networks,'' \emph{IEEE Transactions on Communications},
  vol.~14, no.~8, pp. 1--12, Mar. 2017.

\bibitem{WPTBeam16}
R.~Du, C.~Fischione, and M.~Xiao, ``Lifetime maximization for sensor networks
  with wireless energy transfer,'' in \emph{IEEE ICC}, Kuala Lumpur, Malaysia,
  May 2016.

\bibitem{StocProgBook}
D.~Birge and F.~Louveaux, \emph{Introduction to Stochastic Programming}.\hskip
  1em plus 0.5em minus 0.4em\relax Springer, 1997.

\bibitem{RFPowersim}
``Radio transmit power,''
  \url{https://www.cisco.com/c/en/us/td/docs/routers/access/wireless/software/guide/RadioTransmitPower.html}.

\bibitem{sim1}
``Waspmote datasheet,''
  \url{http://www.libelium.com/downloads/documentation/waspmote_datasheet.pdf}.

\end{thebibliography}

\end{document}